\documentstyle[12pt]{article}

\input epsf.tex

\renewcommand{\theequation}{\thesection.\arabic{equation}}
\newcommand{\be}{\begin{equation}}   \newcommand{\ee}{\end{equation}}
\newcommand{\bear}{\begin{eqnarray}}
\newcommand{\eear}{\end{eqnarray}}
\newcommand{\ba}{\begin{array}}      \newcommand{\ea}{\end{array}}

\newcommand{\gae}{\begin{array}{c}\,\sim\vspace{-21pt}\\> \end{array}}
\begin{document}
\pagestyle{empty}
\begin{titlepage}

\vspace*{-8mm}
\noindent 
\makebox[11.5cm][l]{MIT-CTP-2555} hep-ph/9607487 \\
\makebox[11.5cm][l]{BUHEP-96-25} July 30, 1996\\
\makebox[11.5cm][l]{Submitted to Nucl.\ Phys.\ {\bf B}}
Revised September 4, 1996

\vspace{2.cm}
\begin{center}
  {\LARGE {\bf  Vacuum Instability in Low-Energy \\ [2mm] 
                Supersymmetry Breaking Models}}\\
\vspace{42pt}
Indranil Dasgupta $^{a,}$\footnote{e-mail address:
dgupta@budoe.bu.edu},
Bogdan A. Dobrescu $^{a,}$\footnote{e-mail address:
dobrescu@budoe.bu.edu}
and Lisa Randall $^{b,}$\footnote{e-mail address:
lisa@ctptop.mit.edu}

\vspace*{0.5cm}

{\it a) \ \ Department of Physics, Boston University \\
590 Commonwealth Avenue, Boston, MA 02215, USA}

\vspace*{0.5cm}
{\it b) \ \ Laboratory for Nuclear Science and Department of 
Physics\\ 
Massachusetts Institute of Technology \\ Cambridge, MA  02139}

\vskip 2.4cm
\end{center}
\baselineskip=18pt

\begin{abstract}

{\normalsize

We show that for the simplest models of gauge mediated supersymmetry
breaking, including all existing models, the true vacuum will not 
preserve QCD if it communicates supersymmetry breaking to 
the visible sector. We show that the desired supersymmetry 
breaking color preserving vacuum can nonetheless be stable 
cosmologically, but only if certain couplings are sufficiently small.  
We also present simple modifications to the sector which communicates 
supersymmetry breaking in which the true vacuum is acceptable, 
suggesting desirable properties to be sought in dynamical models of 
supersymmetry breaking.
}

\end{abstract}

\vfill
\end{titlepage}

\baselineskip=18pt  
\pagestyle{plain}
\setcounter{page}{1}

%%%%%%%%%%%%%%%%%%%%%%%%%%%%%%%%%%%%%%%%%%%%%%%%%%%%%%%%%%%%%%%%
\section{Introduction}

Dynamical supersymmetry breaking (DSB) offers a natural 
explanation for the hierarchy between the electroweak scale 
and the Planck scale. In this context, the ``DSB sector''
contains the gauge interactions responsible for supersymmetry
breaking and the chiral superfields charged 
under this group, while the visible sector includes the supersymmetric
standard model. There are several known models which break
supersymmetry dynamically \cite{ads1,dnns,dsb,su1}, and can be used as the
DSB sector. But a model of supersymmetry breaking is not sufficient;
the supersymmetry breaking must be communicated to the visible sector.
At present, there are two scenarios.
Supersymmetry breaking can be transmitted from the DSB sector
to the visible sector either by supergravity or by gauge interactions.
The latter scenario has the virtue of ensuring a natural
suppression of flavor-changing neutral currents \cite{dn}.
Also, if supersymmetry breaking is gauge mediated, the soft breaking
parameters are potentially  determined in terms of a few coupling 
constants.

However, it is difficult to make a phenomenologically successful
model of this type. It is essential to give the gauginos a mass.
Current models \cite{dnns,dns} 
do so by first breaking supersymmetry in one sector,
then generating a VEV for the $F$ component of a singlet field,
coupling this singlet field to messenger quarks, leading to a one
loop gluino
mass of order of the electroweak scale. This is not very
economical,
and ideally, one would like to construct a model in which
the communication of supersymmetry breaking is more direct.
However, constructing more directly connected models can prove
to be difficult.

Were one to try to introduce the messenger quarks into the 
DSB sector, asymptotic freedom would typically
be lost \cite{ads1}. One could try to eliminate the scalar
in favor of a fermion bilinear but that seems not to fit
well into models of DSB so far constructed \cite{gluino}. One might
introduce the singlet directly into the DSB sector \cite{su1,su}. But
it is not clear that even a locally stable minimum 
can be constructed which breaks supersymmetry and preserves
color.

The phenomenology of gauge mediated supersymmetry breaking 
is the subject of several recent studies \cite{phen}.  In this paper,
we explore another question, namely the stability of
phenomenologically successful minima.
Our results regarding the properties of the global minima in 
the simplest known models of this type are presented in the next
section.
In section 3 we consider the possibility that we live in the
false vacuum. More complicated models are discussed in section 4.
A summary can be found in section 5.

%%%%%%%%%%%%%%%%%%%%%%%%%%%%%%%%%%%%%%%%%%%%%%%%%%%%%%%%%%%%%%%%%%%%%%
\section{The Vacua of the Messenger Sector}
\label{sec:color}
\setcounter{equation}{0}

In the existing models of gauge mediated supersymmetry breaking
there is a sector which breaks supersymmetry dynamically
at a scale $\Lambda_s \sim 10^2 - 10^4$ TeV. This DSB sector has
a global symmetry $G_m$ which is weakly gauged so that 
it may be considered a small perturbation to the DSB mechanism.
Supersymmetry breaking is transmitted from the DSB sector to the
visible sector through a ``messenger sector'', which 
consists of a gauge singlet,
$S$, a pair of messenger quarks, $q$ and $\bar{q}$, in the 3 and
$\bar{3}$ of the color SU$(3)_C$ group, 
and a pair of chiral superfields,
$N$ and $P$, in vector-like representations of $G_m$.
If there is a $(G_m)^3$ anomaly in the DSB sector, the messenger sector
should also include other superfields, $E_i$, to cancel the anomaly.
The superpotential of the messenger sector includes the most general
dimension-3 operators:
\be
W_{\rm mes} = 
\kappa S \bar{q}q + \frac{\lambda}{3}S^3 + \lambda_1 P N S
+ W_1(P,N,E_i)~.
\label{sup}
\ee
The $P N S$ term communicates supersymmetry breaking from the 
fields charged under the messenger group to the gauge singlet,
and the $S \bar{q}q$ term links the gauge singlet with the 
messenger quarks. Couplings of $P$ and $N$ to the other 
charged chiral superfields are contained in $W_1$.   
A discrete or $R$ symmetry is required to forbid linear 
or quadratic terms in $S$ and a $\bar{q}q$ term,
which otherwise would have coefficients
of order the Planck scale. We choose a phase definition of the fields
such that the coupling constants,
$\kappa, \lambda, \lambda_1$, are positive.

We focus on the case $G_m \equiv $ U(1).
Let $g$ be the U(1) gauge coupling in the normalization where
$P$, $N$ and $E_i$ have charges +1, $-1$ and $y_i$, respectively. 
Upon integrating out the DSB sector, the only presence of
supersymmetry breaking in the messenger sector is through
an effective potential for the charged scalars \cite{dns}:
\be
V_{\rm SB} = M_1^2 \left(|P|^2 - |N|^2 + y_i|E_i|^2\right)
       + M_2^2 \left(|P|^2 + |N|^2 + y_i^2|E_i|^2\right) + ...
\label{rad}
\ee
where the mass parameters $M_k$, $k = 1, 2$ are related to the 
supersymmetry breaking scale by
\be
M_k^2 = c_k \Lambda_s^2 \left(\frac{g^2}{(4\pi)^2}\right)^k ~.
\ee
Here $c_k$ ($j$ = 1,2) are coefficients
of order unity (in explicit calculations they can be larger,
but perturbation breaks down if they are too large)
with a sign dependent on the content of the DSB sector.
The two terms shown in (\ref{rad}) give masses to the charged scalars
and are due to an one-loop Fayet-Iliopoulos term and 
a two-loop contribution, respectively.
The ellipsis in (\ref{rad}) stand for higher dimensional terms and
contributions from more loops.

The scalar potential of the messenger sector includes also 
a contribution from the U(1) $D$-term,
\be
V_D = \frac{g^2}{2}\left(|P|^2 - |N|^2 + y_i|E_i|^2\right)^2 ~,
\label{dterm}
\ee
and $F$ terms:
\be
V_F = |F_S|^2 + 
\kappa^2 |S|^2 \left(|q|^2 + |\bar{q}|^2\right)
+ \left|\lambda_1 N S + \frac{\partial W_1}{\partial P}\right|^2
+ \left|\lambda_1 P S + \frac{\partial W_1}{\partial N}\right|^2
+ \sum_i \left|\frac{\partial W_1}{\partial E_i}\right|^2 ~,
\label{pot}
\ee
where 
\be
F_S = \kappa \bar{q} q + \lambda S^2 + \lambda_1 P N~.
\ee
The SU(3)$_C$ $D$ term contribution to the 
classical potential vanishes if $q = \bar{q}$.

The desired minimum of the scalar potential of the messenger sector,
\be
V_{\rm mes} = 
V_D\left(|P|^2,|N|^2,|E_i|^2\right) + 
V_{\rm SB}\left(|P|^2,|N|^2,|E_i|^2\right) +
V_F\left(P,N,S,q,\bar{q},E_i\right) ~,
\label{vmes}
\ee
is at $S = \langle S \rangle, 
F_S = \langle F_S\rangle$ and $q = \bar{q} = 0$,
 where the fermion components of $q$ and $\bar{q}$  
have a Dirac mass $\kappa |\langle S \rangle|$, and  
the eigenvalues of the mass squared matrix of the scalar components
 is  $\kappa^2|\langle S\rangle|^2 \pm
\kappa|\langle F_S\rangle|$.
As a result, if $\langle F_S\rangle$ is non-zero, a one loop
gluino mass and two-loop squark masses are produced, leading to
highly predictive models.

However, the $S\bar{q}q$ term in the superpotential is dangerous
because it tends to induce VEVs for $q$ and $\bar{q}$ 
in order to  minimize the $|F_S|^2$ term in the scalar potential.
It is  important to determine whether $q$ and $\bar{q}$ 
actually vanish at the global minimum of $V_{\rm mes}$.
We first analyze this problem in the simplest known  models of this
type \cite{dnns}.

%%%%%%%%%%%%%%%%%%%%%%%%%%%%%%%%%%%%%%%%%%%%%
\subsection{Minimal Models}

We start with the class of minimal models, defined by
\be
\frac{\partial W_1}{\partial P} = \frac{\partial W_1}{\partial N} 
= 0 ~.
\label{fit}
\ee
This includes models without the ``extra'' fields $E$  
(i.e. the fields in the messenger sector which cancel the
potential anomaly of the messenger group from the DSB sector), 
models with $W_1 = 0$ \cite{dnns},
and models with $E$ fields which do not couple to $P$ or $N$.
An examination of eqs.~(\ref{dterm})-(\ref{vmes}) 
shows that the global minimum of 
$V_{\rm mes}$ is at $S = 0$ and $\bar{q}q = -
\frac{\lambda_1}{\kappa}P N$, so that $F_S = 0$.
Therefore, the condition $q = \bar{q} = 0, F_S \neq 0$ could be
realized at most at a {\it local} minimum, i.e. any
vacuum that preserves color and breaks supersymmetry
in the visible sector is unstable. 
Note that this proof does not refer to the actual expression 
of $V_{\rm SB}$, such that it remains true even if the
messenger group, $G_m$, is non-Abelian.

To decide whether there are viable local minima, one has to 
consider explicit models. 
The model presented in \cite{dnns} is the only example 
of this type constructed so far. The messenger U(1) has no anomaly
in the DSB sector, and so there is no $E$ field.
This model, based on a SU(6)$\times$U(1)
supercolor gauge group, is constructed such that the
Fayet-Iliopoulos term vanishes ($M_1 = 0$). 
As a result, there is only one scale
in the messenger sector, $M_2^2 < 0$, produced at two loops, and
one expects that the gauge
singlet VEV can be produced without fine-tuning.
The scalar potential of the messenger sector is given by 
\bear
V_{\rm mes} & = & \frac{g^2}{2} \left(|P|^2 - |N|^2 \right)^2 +
 \left(M_2^2 + \lambda_1^2 |S|^2 \right) \left(|P|^2+|N|^2 \right) +
\kappa^2 |S|^2 \left(|q|^2 + |\bar{q}|^2\right)
\nonumber \\
& + & \left|\kappa \bar{q} q + \lambda S^2 + \lambda_1 P N \right|^2 
\, .
\label{bouncepot}
\eear
This potential is unbounded from below, but higher
order terms in $V_{\rm SB}$ (see eq.~(\ref{rad}))
will rectify this, resulting in a deep global
minimum far away in field space.
As shown below eq.~(\ref{fit}), in this true vacuum the visible sector
is supersymmetric.

To search for viable local minima, we set $q = \bar{q} = 0$
in (\ref{bouncepot}). The remaining potential has only quadratic and
quartic terms in $P$, $N$ and $S$, so that the vacuum manifold 
can be found immediately. At the minimum $S \neq 0$ only if 
\be
\lambda > \lambda_1 ~.
\label{lambda}
\ee
These minima are placed at
\bear
& & q = \bar{q} = 0 \, , \;
|P|^2 = |N|^2 = - M_2^2 {\lambda \over
  \lambda_1^3(2-\lambda_1 /\lambda )} \, , \;
\nonumber \\
& & |S|^2 = - M_2^2
{1-\lambda_1/\lambda \over \lambda_1^2(2-\lambda_1/\lambda )} \, , \;
{\rm Arg} (P N S^{*2}) = - 1 ~.
\label{points}
\eear
and the condition of stability is (see the Appendix)
\be
\lambda_1^3 \le 2\lambda g^2 ~.
\label{plane}
\ee

These are also local minima of $V_{\rm mes}$ only if the following
conditions are satisfied at the points (\ref{points}):
\bear
& & \frac{\partial^2 V_{\rm mes}}{\partial |q|^2} +
\frac{\partial^2 V_{\rm mes}}{\partial |\bar{q}|^2} \ge 0
\label{cond1} \\ [3mm]
& & \frac{\partial^2 V_{\rm mes}}{\partial |q|^2}
\frac{\partial^2 V_{\rm mes}}{\partial |\bar{q}|^2} \ge 
\left( \frac{\partial^2 V_{\rm mes}}{\partial |\bar{q}|
\partial |q|} \right)^2 ~.
\label{cond2}
\eear
Condition (\ref{cond1}) is automatically satisfied, while
condition (\ref{cond2}) requires
\be
\lambda_1 \le \frac{\kappa\lambda}{\kappa + \lambda} ~.
\label{dash}
\ee

In conclusion, the messenger sector of the model \cite{dnns}
has a viable false vacuum (see eq.~(\ref{points})) if and only if
(\ref{lambda}), (\ref{plane}) and (\ref{dash}) are satisfied.
In section 3 we study the stability of this false vacuum.

Note that the 
inclusion of a pair of vector-like leptons which couple to $S$,
for giving mass to the usual sleptons and gauginos
\cite{dns}, does not change the position of the false vacuum.

%%%%%%%%%%%%%%%%%%%%%%%%%%%%%%%%%%%%%%%
\subsection{Models with One $E$ Field}

We have shown so far that in the models without extra fields, $E$,
coupled to $P$ or $N$, the deepest minimum is not viable.
Although  models with extra fields might seem less attractive,
their existence will change the  vacuum structure,
and can conceivably stabilize the preferred minimum.
In the remainder of this section we study
whether it is possible to have a viable true 
vacuum in models where there is only one charged
superfield, $E$, in addition to $P$ and $N$ \cite{dns}.

There are two ways of including $E$ in the 
renormalizable superpotential
($\lambda_2 > 0$):
\be
W_1 = \frac{\lambda_2}{2} E N^2
\label{n2e}
\ee
if $E$ has charge $y = 2$, or 
\be
W_1 = \frac{\lambda_2}{2} E^2 N
\label{e2n}
\ee
if $y = 1/2$. The cases $y = -2$ and $y = -1/2$ can be derived 
from these ones by putting $P \leftrightarrow N, 
\ M^2_1 \leftrightarrow - M^2_1$.

For generic parameters, $V_{\rm mes}$ cannot be minimized analytically.
However, the global minimum can be identified for a large range
of parameters. To this end, we split $V_{\rm mes}$ into two pieces:
\be
V_{\rm mes} = V_a + V_b
\ee
where
\be
V_a \equiv V_D + V_{\rm SB} +
\left|\frac{\partial W_1}{\partial E}\right|^2
\ee
and 
\be
V_b \geq 0 ~.
\ee

When $W_1$ is given by (\ref{n2e}),
$V_b = 0$ if and only if $S = 0, \, N E = 0$ and  
$\bar{q}q = - \frac{\lambda_1}{\kappa}P N$.
Hence, if the global minimum of $V_a$
is at $N E = 0$, then the global minimum of $V_{\rm mes}$
is at $S = 0, \, F_S = 0$.
A straightforward minimization of $V_a$ shows that 
$N E \neq 0$ only if
\be
M_2^2 < 0 
\ee
and
\be
\frac{M_1^2}{|M_2^2|} < 2 + \frac{6 g^2}{\lambda_2^2} ~.
\label{lim}
\ee 

This range of parameters, which might allow supersymmetry 
breaking in the visible sector, can be further reduced.
It would be useful
to see for what range of parameters the minima of $V_{\rm
  mes}(q = \bar{q} = 0)$ are lower than the global minimum
of $V_{\rm mes}(S = 0)$,
because this is a necessary condition for the 
existence of a viable true vacuum.
However, $V_{\rm mes}(q = \bar{q} = 0)$
contains quartic, quadratic and linear terms in $S, P, N$ and $E$,
and cannot be minimized analytically. 
Our strategy is to find a function that bounds $V_{\rm mes}(q =
\bar{q} = 0)$ from below and can be minimized analytically,
and to compare its minimum with
the global minimum of $V_{\rm mes}(S = 0)$. 
Consider 
\be 
V_1 \equiv V_{\rm mes}(q = \bar{q} = 0) - |F_S|^2 ~.
\ee
It is easy to find the minimum of $V_1$ with respect to $S$:
\be
V_1(P,N,E)_{S_{\rm min}}
= V_D + V_{\rm SB} + \frac{\lambda_2^2}{4} |N|^4 +
\lambda_2^2 |N E|^2 -
\frac{\lambda_2^2|N P E|^2}{|N|^2 + |P|^2}~.
\ee
Again, this potential cannot be minimized analytically, so we need
to bound it from below.
For any $N$, $P$ and $E$, 
\be
V_1(P,N,E)_{S_{\rm min}}
\geq V_2 \equiv
V_a + \frac{\lambda_2^2}{4} |E|^2 \left(3|N|^2 - |P|^2\right)~,
\ee
the equality being satisfied if $|N| = |P|$ or $E = 0$.

Now we have a simple potential, $V_2$, which is smaller 
than $V_{\rm mes}(q = \bar{q} = 0)$ at any point in field space,
and we can compare it with $V_{\rm mes}(S = 0)$ at $\bar{q}q = -
\frac{\lambda_1}{\kappa}P N$:
\be 
V_{\rm mes}(S = 0)_{\bar{q}q_{\rm min}} =
V_a + \lambda_2^2 |N E|^2 ~. 
\ee
They are equal at their global minima provided 
\be
|E|^2 \left(|N|^2 + |P|^2\right) = 0 
\label{cond}
\ee
at the global minimum of $V_2$.
If this condition is satisfied, then 
\be
V_{\rm mes}(S = 0)_{\rm min} < 
V_{\rm mes}(q = \bar{q} = 0)_{\rm min}
\ee
whenever $F_S \neq 0$, so that
the global minimum of $V_{\rm mes}$ breaks color or preserves
supersymmetry .

For $\lambda_2^2 < 16 g^2$, $V_2$ is bounded from below and it can be
minimized analytically. When $M_2^2 < 0$, the result is that 
eq.~(\ref{cond}) is not satisfied only if
\be
\frac{M_1^2}{|M_2^2|} \leq \left\{ \begin{array}{rcl} 
1 \; & , & \, {\rm for} \,\; \frac{g^2}{\lambda_2^2} \geq \frac{1}{8} 
\\ [3mm]
2 - \frac{8 g^2}{\lambda_2^2} & , & \, {\rm for} \,\; \frac{1}{8} >
\frac{g^2}{\lambda_2^2} > \frac{1}{16} 
\end{array} \right. 
\ee
For $\lambda_2^2 > 16 g^2$, eq.~(\ref{lim}) requires
\be
\frac{M_1^2}{|M_2^2|} < \frac{19}{8} ~.
\label{ineq}
\ee

These are severe constraints on the DSB sector. For instance,
the model of this type constructed in \cite{dns} gives
\be
\frac{M_1^2}{|M_2^2|} = \frac{\pi^2}{2 g^2} ~.
\label{ratio}
\ee
Thus, a necessary (but not sufficient) condition for 
a stable viable vacuum in this model is
$g > 1.4$ and $\lambda_2 > 5.7$, or $g > 1.8$.
However, for such large values of $g$, the DSB and messenger
sectors cannot be treated separately. One has to minimize
the complete scalar potential and decide whether the global minimum
is well behaved.

Note that whenever the messenger sector has a U$(1)^3$ anomaly, as in
this class of models, cancelled by the DSB sector,
no symmetry can forbid a Fayet-Iliopoulos term.
We expect $|M_1^2| \gae \pi^2 |M_2^2|$, so that 
other models in this class might be viable only if 
they yield a negative $M_1^2$, which is not the case
of the DSB sectors whose spectra have been investigated
\cite{dns}.

Finally, consider the case where 
$W_1$ is given by (\ref{e2n}).
It is easy to see that  $V_b = 0$ if and only if 
$S = 0$, $\bar{q}q = - \frac{\lambda_1}{\kappa}P N$ and $E = 0$.
For $M_2^2 \geq 0$, $V_a$ is minimum at $E = 0$.
For $M_2^2 < 0$, ignoring the higher order terms
in $V_{\rm SB}$ (the ellipsis in eq.(\ref{rad})) we find:
$V_a \rightarrow -\infty$ 
for $|N| \rightarrow \infty$ and $|E|^2 < - 2 M_2^2/\lambda_2^2$,
but $V_a(E = 0) < V_a(E \neq 0)$.
Thus, if the perturbative expansion (\ref{rad}) is valid, 
then $V_a$ and
$V_b$ are minimum at $E = S = F_S = 0$, so that supersymmetry is
unbroken in the low energy sector.

%%%%%%%%%%%%%%%%%%%%%%%%%%%%%%%%%%%%%%%%%%%%%%%%%%%%%%%%%%%%%%%%%%%%%
\section {Vacuum Tunnelling}
\label{sec:tunnel}
\setcounter{equation}{0}

So far we have shown that for a large class of models, including all 
the models existing in the literature, there are vacua with viable
properties for some range of parameters, but they are unstable. 
Thus, these models can describe   nature only if: \\
i) the early universe was placed in the false vacuum by a phase
transition;\\
ii) thermal fluctuations did not cause a transition at a later time
from the false vacuum to the true one;\\
iii) the lifetime of the false vacuum is at least of the order
of the present age of the universe.\\
Here we study only the restrictions imposed by condition iii).

A false vacuum will eventually decay to the true vacuum by
tunnelling, but its lifetime could be arbitrarily long if 
some coupling constants are fine-tuned. 
However, if condition iii) restricts the parameter space
too much, then these models become unsatisfactory because
the goal of DSB is to avoid fine-tuning. 

The zero temperature vacuum tunnelling rate may be computed
semi-classically \cite {coleman}.
The transition probability per
unit volume per unit time is given by:
\be
{\Gamma \over V}= Ae^{-S_E[ \overline{\phi} ]} \; ,
\label {bounce}
\ee
where $S_E[ \overline{\phi} ]$ is the Euclidean action for the
``bounce'' configuration $\overline {\phi}$. $A$ has the dimensions of
(mass)$^4$ and is of the order of $(10 \, \rm {TeV})^4$ in the present
case.
The space-time four-volume available for the transition is about $t^4$
where $t \approx 10^{10}$ years is the age of the universe. For the
stability of the false vacuum it is reasonable to demand that
$t^4{\Gamma / V} \le 1$, which implies
\be
S_E[ \overline{\phi} ] \ge 400 ~.
\label{value}
\ee

A reliable estimate of the bounce action can be obtained only
numerically in most cases (see \cite {coleman} for an analytical
approximation in the ``thin wall'' case). This computation
is simple if there is {\it only one} scalar field.
The bounce
solution is an $O(4)$ invariant non-trivial field configuration
which is an extremum of the Euclidian action, and obeys the boundary
conditions:
$(d\overline{\phi}(r)/dr)_{r=0} = 0 \, , \;
\overline {\phi}(r \to \infty) = \phi ^{f}$,
where $\phi$ is the scalar field and $\phi ^{f}$ is its value at the
false vacuum. In this case one can solve the Euclidian equations of
motion from some initial point $\phi (0)$ and look at the limiting
value
$\phi (\infty)$. For arbitrary values of $\phi (0)$ the value of $\phi
(\infty)$ is either an ``overshoot'' ($\phi (\infty)>\phi^f$) or an
``undershoot'' ($\phi (\infty)<\phi^f$). Since the correct value of
$\phi (0)$ (called the ``escape point'')
must lie between two trial values which end in an overshoot and an
undershoot, the search converges rapidly by bisections.

When two or more scalar fields are involved,
the bracketing property of the overshoot and the undershoot is
lost making the search for the bounce a much more difficult task.
It is however possible to find an upper bound on the bounce action
by replacing the multi-field potential by a suitably chosen single
field
potential.

We apply this procedure to the minimal model \cite{dnns}, whose
viable false vacuum has the coordinates given by eq.~(\ref{points}), by
 choosing a convenient direction in the
  field space and replacing the multidimensional space by a one
dimensional space extending in that direction and passing through the
  false vacuum. Then we find the escape point in this one dimensional
case by bisections of undershoots and overshoots. The
solution obtained by this method is not a bounce solution of the
multidimensional case, but it gives
an upper bound on the least bounce action. We have
further refined our bound on the least bounce action by repeating
the above procedure for different directions in field space to find
the
direction that gives the least bounce action.
Finally we search
for the bounce in the full multidimensional space in a neighborhood of
the straight line trajectory in the ``best'' direction. The last step
improves the upper bound set on $\lambda_1$ by a factor of $1/2$.
When searching for
the bounce in the multidimensional space, we use methods similar to
ref. \cite{klg}.

The one dimensional cross section in the field space that we finally
chose as the initial guess for the trajectory for the multidimensional
search of the bounce connects the false vacuum (\ref{points})
to the point with $|P| = |N|$ unchanged and
\be
S = 0\, , \;\; |q|^2 = |\bar{q}|^2 = -M_2^2 {\lambda \over
\kappa \lambda_1^2 (2 -\lambda_1 /\lambda )} ~.
\ee
It can be verified that the potential at the latter
point is less than the potential at the false vacuum, hence the one
dimensional cross section has a bounce solution.

Our numerical results are shown in
Figure 1. We have searched for the bounce for values
of $\lambda , \lambda_1$ and $\kappa$ ranging from 0.1 to
1. The
graphs are plotted for $\kappa =0.4$ and $\kappa=1.0$.
For a given value of $\kappa$, the region above the dashed curve in
Figure 1 is ruled out analytically because there is no local minimum
(see eq.~(\ref{dash})), while
%$\lambda_1 > \kappa\lambda/(\kappa + 2\lambda)$
the region above the solid line is ruled out by our numerical results
(the
bounce action is less
than $400$). From the graphs one can see that in this
model the false vacuum can not be long lived when $\lambda_1 > 0.1$.
Note that these bounds are independent of $g$. 

%%%%%% here is inserted the figure (the ps file is called dsb_fig.ps)
\newpage\vspace*{-1.3cm}
\vspace*{-1.3cm}
\centerline{\epsfxsize=3.0in\epsfbox{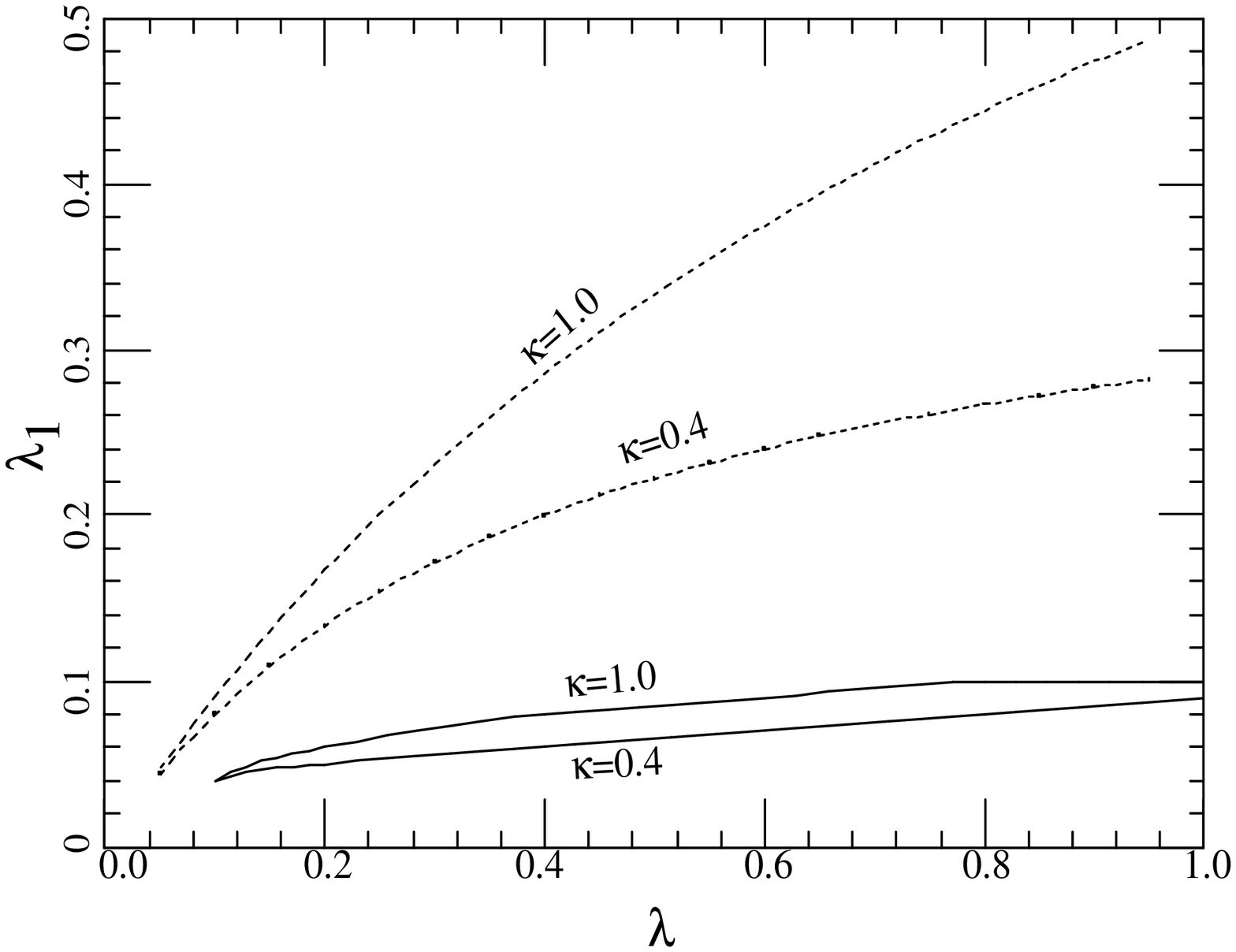}}
\vspace{-1.5cm}
%%%%%% caption begins %%%%%%
\makebox[0.8in][l]{\hspace{2ex} Fig. 1.}
\parbox[t]{4.8in}{ {\small Vacuum Stability.
The region above the solid (dashed) lines is ruled out
numerically (analytically). } }
\smallskip
%%%%%% caption ends %%%%%%

\vspace{4mm}
In conclusion, the numerical analysis of the vacuum tunnelling
indicates that in the interesting range of parameters where the theory
can be treated pertubatively, at least one coupling is less than
$0.1$. This may not be regarded as a small enough number to qualify
the
theory as unnatural. Thus, viable and long lived false vacuua in this
theory  are not   ruled out at this stage. However, a more robust
numerical analysis  might constrain the parameter space even further.

%%%%%%%%%%%%%%%%%%%%%%%%%%%%%%%%%%%%%%%%%%%%%%%%%%%%%%%%%%%%%%%%%%%%%
%%%%%%%%%%%%%%%%%%%%%%%%%%%%%%%%%%%%%%%%%%%%%%%%%%%%%%%%%%%%%%%%%%%%%
%%%%%%%%%%%%%%%%%%%%%%%%%%%%%%%%%%%%%%%%%%%%%%%%%%%%%%%%%%%%%%%%%%%%%
\section {Enlarging the Messenger Sector}
\label{sec:more}
\setcounter{equation}{0}

Although it is conceivable that the universe is at present in a
false vacuum, 
it would be preferable if the messenger sector had viable stable vacua.

If new fields are added to the messenger sector, then there will be
more $F$-type terms contributing to the scalar potential, so that the
position of the true vacuum is likely to change. One may hope that in
some cases the true vacuum is viable.
Broadly speaking, the minimal messenger sector can be
modified in three ways:

\noindent
(i) modifying/enlarging the matter content
transforming under  SU$(3)_C$

\noindent
(ii) enlarging the sector transforming non-trivially under
the messenger group $G_m$ ($P,N,E$ etc.),

\noindent
(iii) enlarging the singlet sector ($S$).

In the following we 
consider modifications of only one of the above kinds at a
time. Some of our results can be extended in a straightforward manner
to the cases when several parts of the messenger sector are changed at
once.

\vspace{2mm}
{\it (i) SU$(3)_C$ colored sector.} 

The most severe constraint on this sector is
the preservation of asymptotic freedom for QCD. In supersymmetric QCD,
apart from the chiral multiplets corresponding to the usual quarks,
there is still room for either
up to three $(3+ \overline 3)$s or one 8. 
With this chiral content the results of section 2 will not change.

\vspace{2mm}
{\it (ii) The $G_m$ colored sector.} 

We have not explored the many
possibilities when $G_m$ is non-Abelian. 
Note, however, that our result regarding the true vacuum in minimal
models (see section 2.1) is true for any $G_m$
if there are no new terms in the superpotential

In the simplest case, when $G_m$
is a U(1), one can enlarge the minimal messenger sector by introducing
additional charged fields. Although the potential is hard to minimize
in the general case, considerable simplification is achieved by taking
some of the coupling constants to be small. For instance, consider
enlarging
this sector by introducing two fields $E_1$ and $E_2$ with charges 
$2$ and $-{1 \over 2}$ respectively. The superpotential is given by
(\ref{sup}) with,
\be 
W_1(P,N,E_i) = {\lambda_2 \over 2} N^2E_1 + {\lambda_3 \over 2}
PE_2^2 ~.
\label {onesinglet}
\ee
When $\lambda_1 = 0$ it is possible to have a global
minimum of the potential at $S = q =
\overline q = E_1 =0$ and $P, N, E_2 \ne 0$. In particular, this
happens when
\be
1 + \frac{4 g^2}{\lambda_2^2} - \frac{3 g^2}{4 \lambda_3^2}
< - \frac{M_1^2}{M_2^2} < 1 + \frac{4 g^2}{\lambda_2^2} ~,
\ee
with $M_2^2 < 0$ and $M_1^2 > 0$. 
When $\lambda_1$ is given a small positive value,
this minimum is shifted.
One can parametrize the small changes in the coordinates
of the minimum by  $\delta \phi = l_{\phi} x$, where $\phi$ stands for
$P, N, E_1, E_2, S$ and $q$, $l_{\phi}$ are complex numbers, and $x >
0$.
The change in the potential is a quartic polynomial in $x$ and one
can optimize the choice $l_{\phi}$ in the limit of $\lambda_1
\rightarrow 0$. We have found that the optimal choice for $l_q$ is
$0$,
i.e the minimum can be perturbed to move to $q =\overline q =
0$ and $P, N, E_1, E_2, S, F_S \ne 0$, provided $\lambda_3 $ and $
\lambda_1 /
\lambda_3 ^2$ are small and $\kappa > \lambda $.

One can also search for parameters with a good vacuum in which
$\lambda_1$ is not small.  For example, we find that $\lambda_1=0.3,\,
\lambda_2=1.0,\, \lambda_3=0.8,\,
\lambda=1.0,\, \kappa=2.0$, $g=1.0$ and $|{M_1/M_2}|=10$
yields a viable vacuum at $q = \bar{q} = 0$,
$|S| = 2.89,\, 
|N| = 7.45,\, |P| = 0.221,\, |E_1| = 0.014,\, |E_2| = 6.21$,
in units of $|M_2|$, and $|F_S| = 6.19 |M_2|^2$. 

This example demonstrates that color symmetric vacua may be
obtained in the messenger sector if there are at least two $E$ fields.
It remains a challenge to construct a DSB model
in which anomalies cancel and the ratio $|M_1/M_2|$
is suitable.

\vspace{2mm}
{\it (iii) The singlet sector.} 

Consider now a messenger sector with two gauge singlets, $S$ and
$S^{\prime}$.
The most general superpotential, up to transformations which leave 
the minimal Kahler potential invariant, includes the coupling of only
one singlet to $PN$, and the couplings of the both singlets
to $\bar{q}q$:
\be
W = (\kappa S + \kappa^{\prime} S^{\prime}) \bar{q}q 
+ \frac{\lambda}{3}S^3
+ \frac{\lambda ^{\prime}}{3}S^{\prime 3}  
+ \lambda_0 S^2 S^{\prime}
+ \lambda_0^{\prime} S S^{\prime 2}
+ \lambda_1 S P N 
+ W_1(P,N,E_i)~.
\label{twosinglet}
\ee

We will consider the case where there is one $E$ 
field\footnote{The two singlet model with no E field is 
  discussed in ref.\cite{new}.}
in order to use the simplest DSB sector, i.e. the SU(3) $\times$ SU(2)
model, as described in \cite{dns}.

The case $W_1 = 0$ (which corresponds to a messenger charge of $E$, $y$,
different than $\pm 2, \pm 1/2$)
can be analyzed as in the one singlet models
(see section 2.1).
The global minimum is at $S = S^{\prime} = 0$, $F_S = F_{S^{\prime}} =
0$, so that supersymmetry is unbroken in the visible sector.

The case $W_1 = {\lambda_2 \over 2} E N^2$ (with a single $E$ field)
is already complicated. There are 8 coupling constants in the
superpotential, and only 5 phases can be absorbed by field
transformations. Thus, together with $g$ there are 12 parameters
that can be varied when one searches for a minimum of the
11 dimensional field space.
Note that the DSB sector remains the same as in ref.~\cite {dns},
so that the one- and two- loop masses of the charged scalars
satisfy eq.~(\ref{ratio}).

Analytically, using the method described in the case of the model 
with two $E$ fields,
we have found that close to the point $\lambda_1 =
\lambda_2 = 0$ in the parameter space it is
possible to obtain a global minimum of the potential at $q, \overline
q=0$ and $S, S^{\prime}, P, N, F_{S} \ne 0$.

We have numerically minimized the potential and have found that, for 
a range of parameters, the deepest minimum is viable.
For example, when 
\bear
& g=0.8\, , \;\; \lambda_1=0.3\, , \;\;  \lambda_2=0.2\, , \;\;
\lambda=1.0\, , \;\; \lambda^{\prime}=0.5 & \nonumber\\
& \lambda_0=0.25\, , \;\;  \lambda_0^{\prime}=0.5\, , \;\;  k=1.0
\, , \;\;  k^{\prime}=2.0 ~, &
\label{par}
\eear
the vacuum is at 
\bear
& q = \bar{q} = 0\, , \;\ S \approx 1.906\, , \;\;  
S^{\prime} \approx - 3.982 & \nonumber\\
& P \approx 3.688\, , \;\; N \approx - 7.792\, , \;\ 
E \approx 4.295 ~, &
\label{minim}
\eear
in units of $|M_2|$. In fact this is a set of degenerate vacua 
connected by phase transformations which preserve the following
relations:
\be {\rm Arg}(S S^{\prime}) = {\rm Arg}(P N S^{*2}) = 
{\rm Arg}(S P N^* E^*) = \pi ~.
\ee
At these minima, $|F_S| \approx 0.856 \, |M_2^2|,\; 
|F_{S^{\prime}}| \approx
1.249 \, |M_2^2|$, and the vacuum energy density of the messenger
sector is 
\be
V_{\rm min} \approx - 113.5 \, M_2^4 ~.
\label{deep}
\ee

In the absence of analytical
constraints it is not possible to be absolutely
sure that a minimum found numerically is the global one.
However, for this model it is possible to find analytically
the minimum when $S = S^{\prime} = 0$. When the parameters are
given by eq.~(\ref{par}), the minimum value of the potential is
\bear
V(S = S^{\prime} = 0)_{{\rm min}} & = & - \, \frac{3 M_2^4}{6 g^2 -
  \lambda_2^2}
\left( \frac{M_1^4}{M_2^4} - 2 \frac{M_1^2}{|M_2|^2} + 6
  \frac{g^2}{\lambda_2^2} \right) \\ [2mm]
& = & - 110.6 \, M_2^4 ~.
\label{shallow}
\eear
Thus, the viable minimum (\ref{minim}), found numerically, is deeper
than the minimum at $S = S^{\prime} = 0$. The values (\ref{deep})
and (\ref{shallow}) are close to each other because the bulk of the
two potentials at the minimum comes from the same $M_1^2 |N|^2$ term. 
Although we have not ruled out analytically a deeper minimum at 
$S,S^{\prime},q \neq 0$, this is unlikely to exist due to the
$F_q$ and $F_{\bar{q}}$ terms which can be large.
 
Note that a similar two singlet model, with an additional 
$S^{\prime} H_u H_d$ coupling of the Higgs superfields, 
was discussed in ref.~\cite{dns} as a source for a $\mu$ term. 
However, in that case the $\lambda_0$
and $\lambda_0^{\prime}$ coupling constants have to be small,
to allow the hierarchy between the electroweak scale and the messenger
quark masses.
It is easy to see that under these conditions both 
$V_{\rm mes}(q = \bar{q} = 0)$ and $V_{\rm mes}(S = 0)$
have the global minimum at $S^{\prime} \approx 0$ and
$F_{S^{\prime}} \approx 0$. Therefore, these two potentials
can be compared exactly as in section 2.2.
The result of this analysis is that the true vacuum of the model 
used to produce a $\mu$ term is not
viable (at least in the case of a weakly gauged messenger group).

%%%%%%%%%%%%%%%%%%%%%%%%%%%%%%%%%%%%%%%%%%%%%%%%%%%%%%%%%%%%%%%
\section{Summary and Conclusions}
\label{sec:conc}
\setcounter{equation}{0}

Many of the naturalness problems of the standard model and its
supersymmetric versions can be solved within models with
dynamical supersymmetry breaking at low energy.
However, the predictive power of this type of models is accompanied
by important phenomenological restrictions.
We have analyzed the properties of the vacua of all the potentially
realistic models with DSB existing in the literature and their simple
extensions and generalizations. 
The main issue is that the messenger quarks should not develop
VEVs while learning of supersymmetry breaking from the gauge singlet
which couples to the $P$ and $N$ fields, which are 
charged under the messenger group.
Here we summarize our main conclusions.

\noindent
{\bf 1.\ } 
In the case of the minimal models, where $P$ and $N$ couple only to
the gauge
singlet (such as the one in \cite{dnns}), the deepest minimum
does not break supersymmetry in the visible sector.
For a range of parameters,
the model of ref.~\cite{dnns} (in which there is no $E$ field and no
Fayet-Iliopoulos term) has a false vacuum with viable properties.
The lifetime of this vacuum is longer than the present age of the
universe provided some coupling constants are sufficiently small.
The condition $\lambda_1 < 0.1$ (see Fig.~1), 
found numerically, is necessary but probably
not sufficient; the actual bound might be much stronger.

\noindent
{\bf 2.\ } In the models where there is only one charged field, $E$,
in addition to $P$ and $N$, the true vacuum
typically breaks color or preserves supersymmetry.
A necessary (but not necessarily sufficient) condition to evade this conclusion
is given by the inequality (\ref{ineq}) involving the one- and two-
loop masses of the charged fields. 
The models of ref.~\cite{dns} do not satisfy this condition unless
the messenger gauge coupling is large, which is in contradiction
with the requirement of a {\it weakly} gauged symmetry of the
DSB sector. It would be useful to build models which have a negative
Fayet-Iliopoulos term such that the necessary condition (\ref{ineq})
is satisfied while keeping weak the messenger gauge interaction.

\noindent
{\bf 3.\ }
Simple extensions of the models with one $E$ field, 
which include additional 
charged or gauge singlet chiral superfields, have viable vacua
for a range of parameters. 
This is the case of a
model with two charged fields in addition to $P$ and $N$.
However, it will be a non-trivial task to find a DSB sector which
cancels the U(1)$^3$ and the mixed U(1) - gravitational anomalies
of the messenger sector for this model.
A model with two gauge singlets and a charged field, $E$,
has a viable true vacuum in the limit where some coupling constants
are small. A numerical search has yielded a viable vacuum for a
natural range of parameters.  This result is encouraging
since the DSB sector is simple, as in ref. \cite {dns}.
However, the second singlet cannot be used to produce a $\mu$ term
at the global minimum.

We conclude by emphasizing the need for extensive studies of the
properties of diverse DSB sectors, and for further searches for 
messenger sectors with viable vacua. It is worth noting that  if
one accepts the possibility of living in a false vacuum,
it suggests new model building alternatives, since the DSB
sector itself might be such that the supersymmetry
breaking vacuum is only  a local minimum.
 Nevertheless, it would be desirable to find viable models in which the
universe is in the true vacuum. 

%%%%%%%%%%%%%%%%%%%%%%%%%%%%%%%%%%%%%%%%%%%%%%%%%%%%%%%%%%%%%%%
\section*{Note}
After the completion of this work we were informed about
ref.~\cite{new} where the minima of the scalar potential
of the messenger sector given in \cite{dnns} are briefly
discussed.

%%%%%%%%%%%%%%%%%%%%%%%%%%%%%%%%%%%%%%%%%%%%%%%%%%%%%%%%%%%%%%%
\section*{Acknowledgements}

We would like to thank Sekhar Chivukula and Erich Poppitz 
for useful conversations. We also thank Csaba Cs\'aki and
Witold Skiba for comments on the manuscript. L.R. thanks 
the Aspen Center for Physics for its hospitality during 
final stages of this project.
We would like to thank Chris Carone for bringing ref.~\cite{new} to
our attention, and for valuble correspondence.

The work of I.D. and B.D. was supported in part by the National Science
Foundation under grant PHY-9057173, and by the Department of Energy
under grant DE-FG02-91ER40676.
The work of L.R. was supported in part by the Department of Energy
under cooperative agreement DE-FC02-94ER40818, NSF
grant PHY89-04035, NSF Young Investigator Award,
Alfred Sloan Foundation Fellowship, and DOE Outstanding Junior 
Investigator Award.

%%%%%%%%%%%%%%%%%%%%%%%%%%%%%%%%%%%%%%%%%%%%%%%%%%%%%%%%%%%%%%%
\section*{Appendix}
\label{sec:app}
\renewcommand{\theequation}{A.\arabic{equation}}
\setcounter{equation}{0}

In this Appendix we derive conditions of stability for the 
color preserving vacua of the minimal model discussed in 
section 2.1.
We seek the minima of $V_{\rm mes}$ (see eq.~(\ref{bouncepot}))
at $q = \bar{q} = 0$ and $S \neq 0$. These satisfy ${\rm Arg} (P N
S^{*2}) = - 1$, so that we need to minimize
a potential which does not depend on complex phases:
\be
V_0\left(|S|,|P|,|N|\right) \equiv V_{\rm mes}
\left(q = \bar{q} = 0, F_S = \lambda |S|^2
- \lambda_1 |P N|\right) ~.
\ee
Its minima are amongst the solutions of
\bear
\!\!\!\ 
\frac{\partial V_0}{\partial |S|} \!& = &\! 2|S|\left\{ 2 \lambda^2
|S|^2 - \lambda_1 \left[2 \lambda |P N| - 
\lambda_1\left(|P|^2 + |N|^2\right)\right]
\right\} = 0
\label{sderiv}\\
\!\!\!\
\frac{\partial V_0}{\partial |P|} \!& = &\! 2|P|
\left[g^2\left(|P|^2  - |N|^2\right)
+ M_2^2 + \lambda_1^2\left(|S|^2+|N|^2\right) \right]
- 2\lambda\lambda_1 |S|^2 |N| = 0
\label{pderiv}\\
\!\!\!
\frac{\partial V_0}{\partial |N|} \!& = &\! 2|N|
\left[g^2\left(|N|^2  - |P|^2\right)
+ M_2^2 + \lambda_1^2\left(|S|^2+|P|^2\right)\right]
- 2\lambda\lambda_1 |S|^2 |P| = 0
\label{nderiv}
\eear
>From (\ref{sderiv}) we see that $S \neq 0$ only if $\lambda >
\lambda_1$. 

If $|P| \neq |N|$, then it can be shown that 
the set of equations (\ref{sderiv})-(\ref{nderiv}) 
has solutions with $|S|$, $|P|$
and $|N|$ positive (for $M_2^2 < 0$) only if
\be
\frac{\lambda_1^3}{2\lambda} \le g^2 < \frac{\lambda_1^2}{2} ~.
\ee
Computing the second derivatives one can see that 
these solutions are not minima but saddle points: 
they are local maxima either in the $|S|-|P|$ or in the $|S|-|N|$
subspaces.

Therefore, we have to solve
(\ref{sderiv})-(\ref{nderiv}) with $|N| = |P|$, and
the result is given  by eq.~(\ref{points}). 
These extrema of $V_0$ are minima provided the matrix
$\partial^2 V_0/(\partial \phi_i \partial\phi_j)$,
with $\phi_i = |S|, |P|, |N|$, has only positive eigenvalues.
These eigenvalues are given by
\bear
\Delta_{1,2} & = &
\frac{1}{2}\left[ A+B+C \pm \sqrt{(A-B-C)^2 + 8 D^2} \right]
\nonumber\\ [1mm]
\Delta_3 & = & C - B
\eear
where $A \equiv \left(\partial^2 V_0/\partial |S|^2\right)_{|N| =
  |P|}$\ ,
$B \equiv \left(\partial^2 V_0/\partial |P|\partial |N|\right)_{|N| =
  |P|}$\ and
\bear
C & \equiv &
\left.\frac{\partial^2 V_0}{\partial |P|^2}\right|_{|N| = |P|} = 
\left.\frac{\partial^2 V_0}{\partial |N|^2}\right|_{|N| = |P|} 
\nonumber\\ [2mm]
D & \equiv &
\left.\frac{\partial^2 V_0}{\partial |S|\partial |P|}\right|_{|N| =
      |P|}
= \left.\frac{\partial^2 V_0}{\partial |S|\partial |N|}\right|_{|N| =
      |P|}
\eear
At the points (\ref{points}) $\Delta_{1,2}$ are always positive, while
$\Delta_3 \ge 0$ if and only if $\lambda_1^3 \le 2\lambda g^2$.
This is the sufficient and necessary
condition\footnote{The condition given in ref.~\cite{new},
$\lambda_1 < g$, is sufficient but not necessary.}
for the extremum of $V_0$ at $|N| = |P|$ to be a local minimum.

%%%%%%%%%%%%%%%%%%%%%%%%%%%%%%%%%%%%%%%%%%%%%%%%%%%%%%%%%%%%%%%

\vfil
\end{document}